\documentclass[12pt]{article}
\pdfoutput=1

\usepackage{graphicx}
\usepackage{dsfont}
\usepackage{amssymb}
\usepackage{amsmath}
\usepackage{mathrsfs}

\begin{document}

\title{Black Hole Lasers Revisited}

\author{U.\ Leonhardt and T.G.\ Philbin \\
School of Physics and Astronomy, \\
University of St.~Andrews, \\
 North Haugh, St.~Andrews, Fife KY16 9SS, Scotland.}

\date{\today}

\maketitle

\begin{abstract}
The production of Hawking radiation by a single horizon is not dependent on the high-frequency dispersion relation of the radiated field. When there are two horizons, however, Corley and Jacobson have shown that superluminal dispersion leads to an amplification of the particle production in the case of bosons. The analytic theory of this "black hole laser" process is quite complicated, so we provide some numerical results in the hope of aiding understanding of this interesting phenomenon. Specifically, we consider sonic horizons in a moving fluid. The theory of elementary excitations in a Bose-Einstein condensate provides an example of  "superluminal" (Bogoliubov) dispersion, so we add Bogoliubov dispersion to Unruh's equation for sound in the fluid.  A white-hole/black-hole horizon pair will then display black hole lasing. Numerical analysis of the evolution of a wave packet gives a clear picture of the amplification process. By utilizing the similarity of a radiating horizon to a parametric amplifier in quantum optics we also analyze the black hole laser as a quantum-optical network.
\end{abstract}


\section{Introduction}
\label{intro}
One important contribution of the theoretical study of black hole analogues has been to help clarify the derivation of the Hawking effect~\cite{unr95,bro95,unr05}. This in turn led to a study~\cite{cor99} of Hawking radiation in a more general context, one that involves, among other features, two horizons. The results of~\cite{cor99} are of particular interest because they offer a scenario, perhaps realizable in a black hole analogue,  in which the Hawking radiation is amplified. Let us begin by describing the background to these ideas.

\begin{figure}
\centering
\includegraphics[height=7cm]{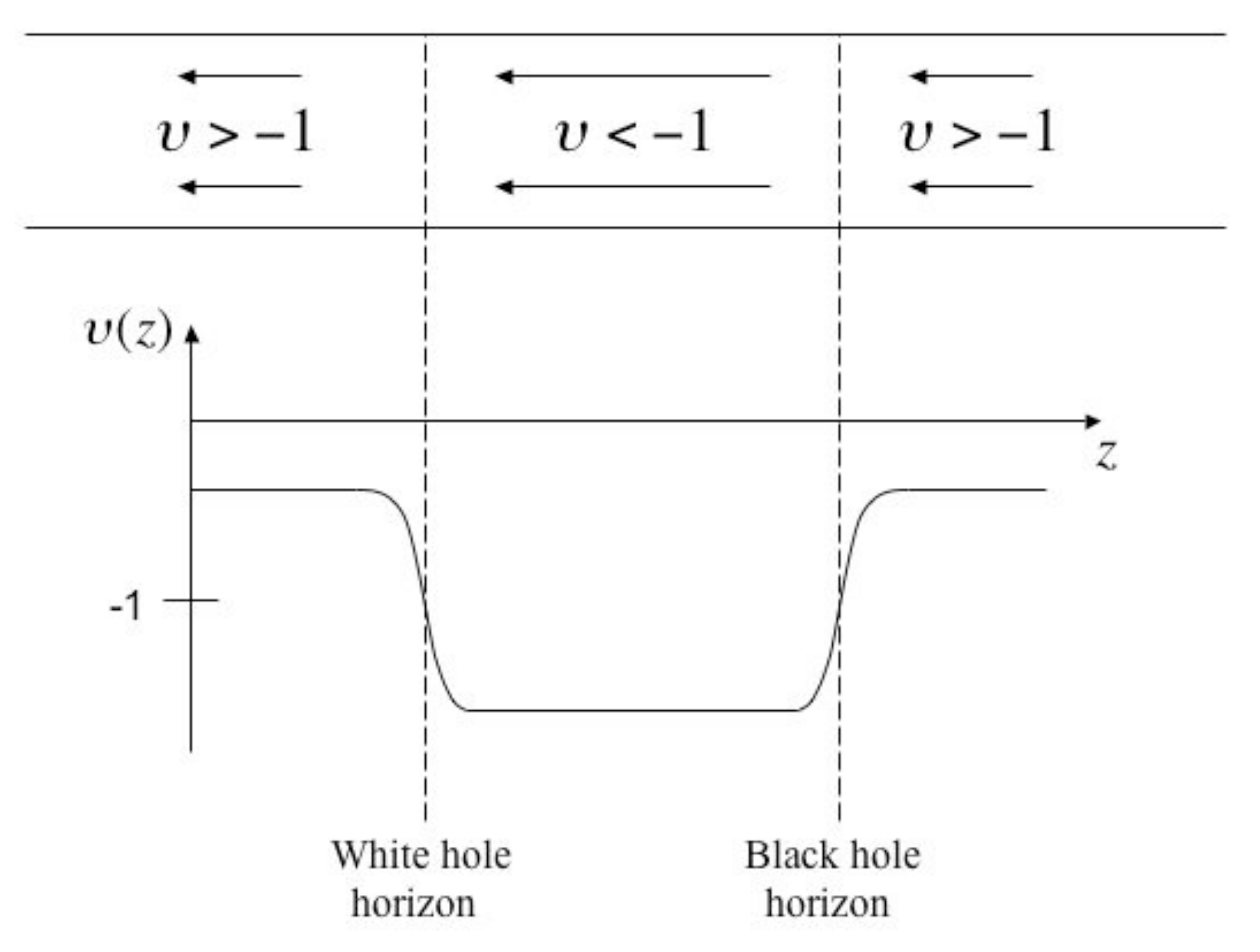}
\caption{Sonic horizons in a moving fluid, in which the speed of sound is $1$. The velocity profile of the fluid, $v(z)$, attains the value $-1$ at two values of $z$; these are horizons for sound waves that are right-moving with respect to the fluid.  At the right-hand horizon right-moving waves are trapped, with waves just to the left of the horizon being swept into the supersonic flow region $v<-1$; no sound can emerge from this region through the horizon, so it is reminiscent of a black hole. At the left-hand horizon right-moving waves become frozen and cannot enter the supersonic flow region; this is reminiscent of a white hole. }
\label{fig1}       
\end{figure}

There is an apparent contradiction in Hawking's semiclassical derivation of black hole evaporation~\cite{haw75},  in that the radiated fields undergo arbitrarily large blue-shifting in the calculation, thus acquiring arbitrarily large masses, which contravenes the underlying assumption that the gravitational effects of the quantum fields may be ignored. This is known as the trans-Plankian problem~\cite{tho85,jac91,unr95,bro95,unr05}.   A similar issue arises in condensed matter analogues such as the sonic black hole~\cite{unr81,unr95}, and to make our discussion concrete from the outset we shall consider sonic horizons in one-dimensional fluid flow (see Fig.~\ref{fig1}). 

When the velocity profile of the fluid is as shown in Fig.~\ref{fig1} two horizons are formed for sound waves that propagate to the right with respect to the fluid. The horizon on the right of the supersonic flow region $v<-1$ behaves like a black hole horizon for right-moving waves, while the horizon on the left of the supersonic flow region behaves like a white hole horizon for these waves. In~\cite{unr81} Unruh showed that in such a system, with some reasonable simplifying assumptions, the equation for a small perturbation $\phi$ of the velocity potential is
\begin{equation} \label{unruh}
(\partial_t+\partial_zv)(\partial_t+v\partial_z)\phi-\partial_z^2\phi=0.
\end{equation}
In terms of a new coordinate $\tau$ defined by
\[
d\tau:=dt+\frac{v}{1-v^2}dz
\]
(\ref{unruh}) is the equation $\phi_{;\mu}^{\ \ ;\mu}=0$ of a scalar field in the black-hole-type metric
\[
ds^2=(1-v^2)d\tau^2-\frac{dz^2}{1-v^2}.
\] 
The results for quantum black holes~\cite{haw75,bir82,bro95b} apply equally well here, so it follows that each horizon will produce a thermal spectrum of phonons with a temperature determined by the quantity that corresponds to the surface gravity at the horizon, namely the absolute value of the slope of the velocity profile:
\begin{equation} \label{temp}
k_BT=\frac{\hbar\alpha}{2\pi}, \qquad \alpha:=\left|\frac{dv}{dz}\right|_{v=-1}.
\end{equation}

\begin{figure}
\centering
\includegraphics[height=7cm]{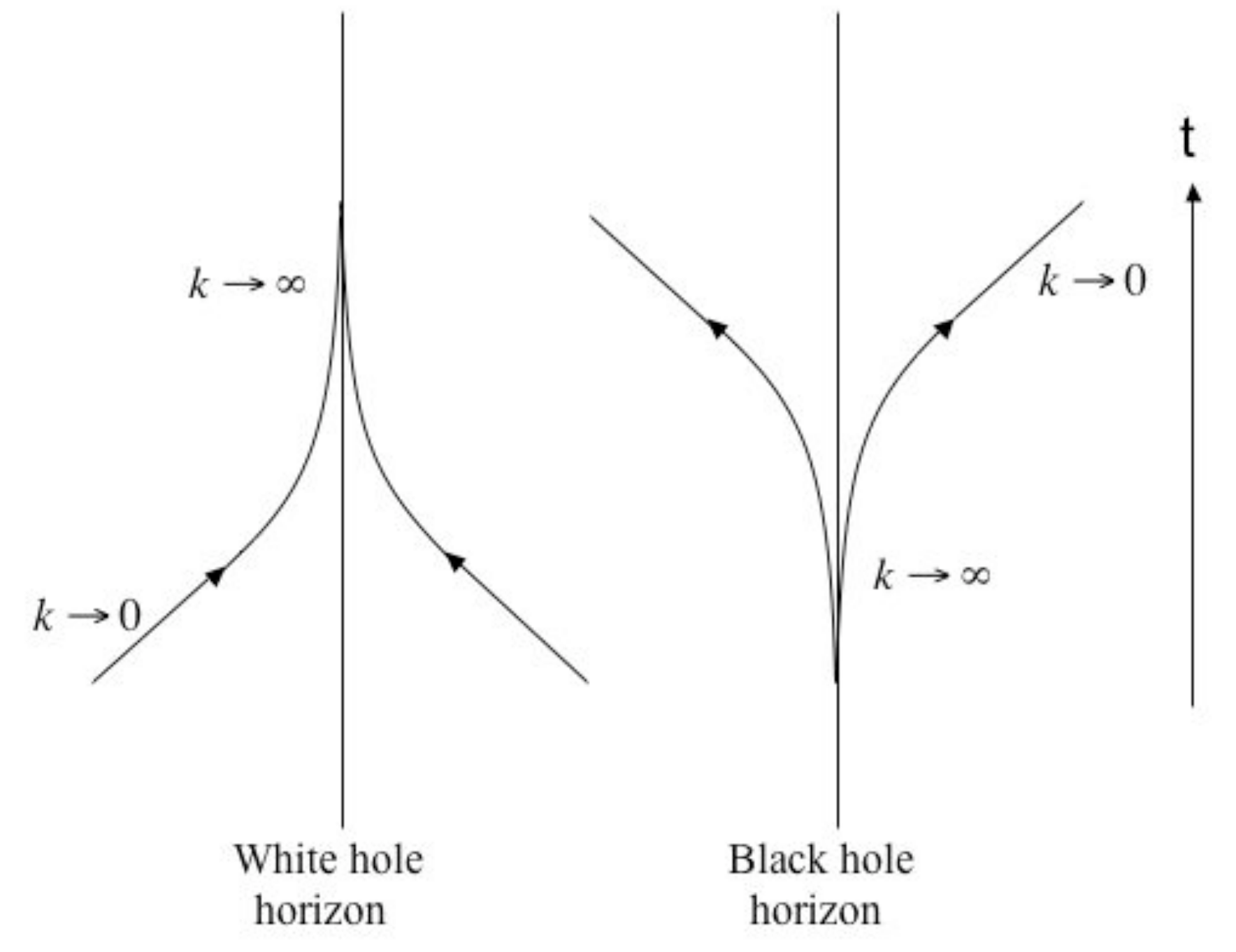}
\caption{Hawking phonons in the fluid flow of Fig.~\ref{fig1}. Real phonons have positive frequency in the fluid-element frame and (\ref{disp1}) shows that for right-moving phonons this frequency ($\omega-vk$) is $\frac{\omega}{1+v}=k$. Thus in the subsonic-flow regions $\omega$ (conserved for each ray) is positive, whereas in the supersonic-flow region it is negative; $k$ is positive for all real phonons. The frequency in the fluid-element frame diverges at the horizons---the trans-Plankian problem.}
\label{fig2}      
\end{figure}

The trajectories  of the created phonons are readily drawn (see Fig.~\ref{fig2}), but they are formally deduced from the dispersion relation of the sound equation (\ref{unruh}). 
Geometrical acoustics applied to (\ref{unruh}) gives the dispersion relation
\begin{equation} \label{disp1}
\omega-vk=\pm k
\end{equation} 
and the Hamilton equations
\begin{gather}
\frac{dz}{dt} =\frac{\partial\omega}{\partial k}=v\pm1, \label{ham1}  \\[6pt] 
\frac{dk}{dt}=-\frac{\partial\omega}{\partial z}=-v'k.  \label{ham2}
\end{gather}
The left-hand side of (\ref{disp1}) is the frequency in the frame co-moving with a fluid element, whereas $\omega$ is the frequency in the laboratory frame; the latter is constant for a time-independent fluid flow (``time-independent Hamiltonian" $d\omega/dt=\partial\omega/\partial t=0)$. Since the Hawking radiation is right-moving with respect to the fluid, we clearly must choose the positive sign in (\ref{disp1}) and hence in (\ref{ham1}) also. By approximating $v(z)$ as a linear function near the horizons we obtain from (\ref{ham1}) and (\ref{ham2}) the ray trajectories of Fig.~\ref{fig2}. The disturbing feature of the rays is the behavior of the wave vector $k$: at the horizons the radiation is exponentially blue-shifted\footnote{We trust the reader will not be too distracted by our mixing of acoustic and optical terminology.}, leading to a diverging frequency in the fluid-element frame. These runaway frequencies are unphysical since (\ref{unruh}) asserts that sound in a fluid element obeys the ordinary wave equation at all wavelengths, in contradiction with the atomic nature of fluids (see Sec.~\ref{disp}). Moreover the conclusion that this Hawking radiation is actually present in the fluid also assumes that (\ref{unruh}) holds at all wavelengths, as exponential blue-shifting of wave packets at the horizon is a feature of the derivation~\cite{haw75,bir82,bro95b}.  Similarly, in the black-hole case the equation used in the calculation~\cite{haw75,bir82,bro95b} does not hold at arbitrarily high frequencies because it ignores the gravity of the fields. For the black hole, a complete resolution of this difficulty will require knowledge of the gravitational physics of quantum fields, i.e.\ quantum gravity, but for the dumb hole the physics is available for a more realistic treatment. The issue to be addressed is the dispersion relation for sound at high frequencies~\cite{unr95}, and a consideration of this will lead us to the black hole laser.

\section{Dispersion}
\label{disp}
\begin{figure}
\centering
\includegraphics[height=1.5cm]{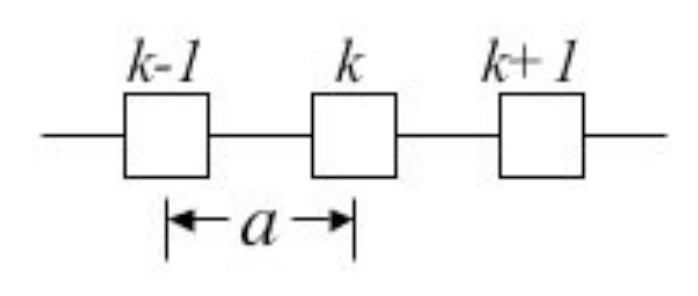}
\caption{A ``classical" picture of fluid atoms. Each atom is labeled and $a$ is the equilibrium separation. }
\label{fig3}     
\end{figure}

In reality one would expect the dispersion relation for sound  to differ from the linear formula (\ref{disp1}) when the wavelength is of the order of the distance between the fluid atoms. What kind of dispersion relation should we expect at these wavelengths? Let us naively picture the fluid atoms as occupying well-defined positions with a separation distance $a$ (see Fig.~\ref{fig3}).
The equation for small oscillations in this chain of atoms is the wave equation with a discretized second-order spatial derivative:
\begin{equation} \label{wavedis}
\partial_t^2\phi_k-\frac{1}{a^2}\left(\phi_{k+1}-2\phi_k+\phi_{k-1}\right)=0.
\end{equation}
A partial differential equation that serves as an approximation to (\ref{wavedis}) is obtained by substituting the following relation, familiar from numerical analysis:
\[
\frac{1}{a^2}\left(\phi_{k+1}-2\phi_k+\phi_{k-1}\right)=\partial_z^2\phi+\frac{a^2}{12}\partial_z^4\phi+O(a^4).
\]
For small $a$ we thus obtain the approximate dispersion relation
\begin{equation} \label{dispn}
\omega^2\approx k^2-\frac{a^2}{12}k^4.
\end{equation}
As the last term in (\ref{dispn}) is negative we find that the group velocity $\frac{d\omega}{dk}$ of waves decreases with  decreasing wavelength, i.e. we have {\it subluminal} or {\it normal} dispersion. 

A nonlinear dispersion relation $\omega=\omega(k)$ such as (\ref{dispn}) limits the amount of blue-shifting that occurs at a horizon, so the embarrassment of infinite frequencies disappears; there then remains the question of whether one can still derive the Hawking effect. Unruh~\cite{unr95} modified the sound equation (\ref{unruh}) to incorporate subluminal dispersion and showed numerically that the prediction of thermal radiation at temperature (\ref{temp}) still holds good, regardless of the details of the high-frequency behavior. Indeed, analysis of generalized models shows that the precise form of the dispersion relation at high frequencies has no effect on the particle production by a single horizon, at least to lowest order~\cite{bro95,unr05}. Unsurprisingly, such a general conclusion involves some assumptions about the short-distance physics at the horizon, and since Planck-scale physics is unknown it remains possible that conditions at this scale conspire to invalidate the evaporation result for real black holes~\cite{unr05}.

We now turn to the main purpose of this Section, namely the introduction of a dispersion relation that, together with the horizon arrangement of Fig.~\ref{fig1}, leads to black hole lasing.

\subsection{Bogoliubov dispersion}
The dispersion relation used by Corley and Jacobson in the original treatment of the black hole laser~\cite{cor99} is well known from the theory of elementary excitations in a Bose--Einstein condensate~\cite{dal99,pit03}.  In this formalism the bosonic atoms are described by an a field  $\hat{\psi}$ but most of the atoms are assumed to form a condensate with macroscopic wave function
\begin{equation}  \label{macro}
\psi_0=\sqrt{\rho_0}e^{iS_0}.
\end{equation}
The deviations of $\hat{\psi}$ from the mean field $\psi_0$ are given by a quantum field $\hat{\phi}$ of small fluctuations:
\[
\hat{\psi}=\psi_0+e^{iS_0}\hat{\phi}.
\]
It is then found that $\hat{\phi}$ satisfies the Bogoliubov--de Gennes equation
\begin{equation} \label{bdg}
i\hbar\partial_t\hat{\phi}=\left(-\frac{\hbar^2}{2m}\nabla^2-\mu+2mc^2\right)\hat{\phi}+mc^2\hat{\phi}^\dagger,
\end{equation}
where $\mu$ is the chemical potential and $c$ is the speed of sound. The field $\hat{\phi}$ is given by the usual mode expansion
\begin{equation} \label{mode}
\hat{\phi}=\sum_\nu\left(u_\nu\hat{a}_\nu+v^*_\nu\hat{a}^\dagger_\nu\right).
\end{equation}
For plane-wave modes
\[
u_\nu=ue^{i\mathbf{k.r}-i\omega t}, \qquad
v_\nu=ve^{i\mathbf{k.r}-i\omega t},
\]
 (\ref{bdg}) and its complex conjugate give
\[
\hbar\omega\begin{pmatrix} u \\ v \\  \end{pmatrix}=
\begin{pmatrix} \frac{\hbar^2k^2}{2m}-\mu+2mc^2 & mc^2 \\
 -mc^2 & -\frac {\hbar^2k^2}{2m}+\mu-2mc^2 \\  \end{pmatrix}
 \begin{pmatrix} u \\ v \\  \end{pmatrix}.
\]
The condition for this matrix equation to have a solution is
\[
\left( \frac{\hbar^2k^2}{2m}-\mu+2mc^2-\hbar\omega\right)\left( -\frac{\hbar^2k^2}{2m}+\mu-2mc^2-\hbar\omega\right)+m^2c^4=0,
\]
and assuming the chemical potential $\mu$ is equal to the local energy of the condensate, $mc^2$, this is the Bogoliubov dispersion relation
\begin{equation} \label{bog}
\omega^2=c^2k^2+\frac{\hbar^2}{4m^2}k^4.
\end{equation}
From (\ref{bog}) we see that the group velocity increases with decreasing wavelength, so it is an example of {\it superluminal} or {\it anomalous} dispersion. Comparison with (\ref{dispn}) shows that the purely quantum term in (\ref{bog}) corresponds to an imaginary separation distance between the atoms, a reflection of the inapplicability of the classical picture of Fig.~\ref{fig3} to a condensate.

Let us now add Bogoliubov dispersion to our model system of Fig.~\ref{fig1}. The appropriate generalization of (\ref{unruh}) is
\begin{equation} \label{unruhmod}
(\partial_t+\partial_zv)(\partial_t+v\partial_z)\phi-\partial_z^2\phi+\frac{1}{k_c^2}\partial_z^4\phi=0,
\end{equation}
where $k_c$ is a constant; this give the dispersion relation
\begin{equation} \label{dispa}
\omega-vk=\pm k\sqrt{1+\frac{k^2}{k_c^2}},
\end{equation}
so that we have Bogoliubov dispersion in a moving fluid. As with (\ref{disp1}) we need only consider the
positive sign in (\ref{dispa}) since our interest is in phonons that propagate to the right with respect to the fluid. 

\begin{figure}
\centering
\includegraphics[height=8cm]{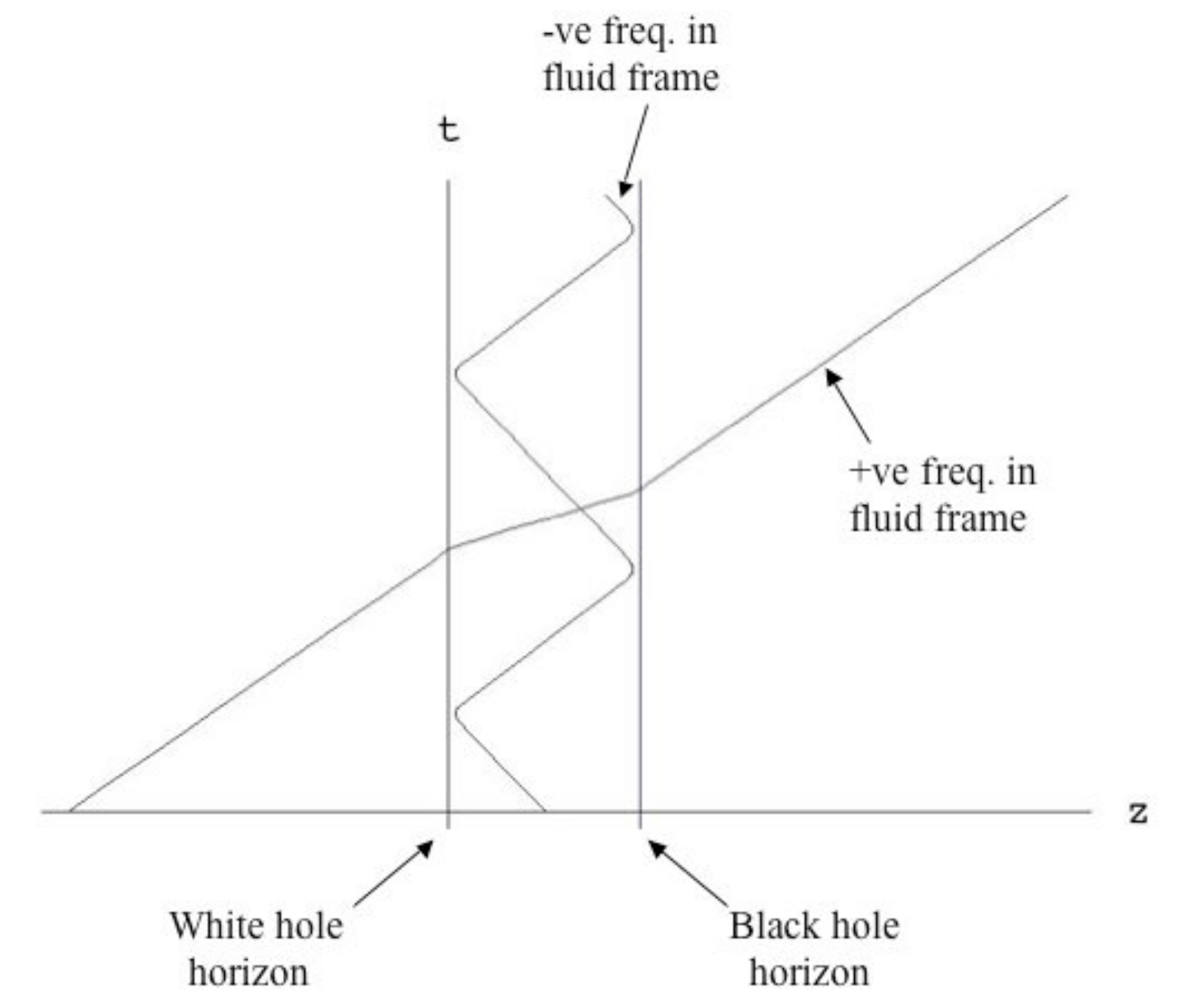}
\caption{Numerical solution of ray trajectories for the dispersion relation (\ref{dispa}). Both rays are right-moving with respect to the fluid and have the same positive value of $\omega$. The ray that is confined to the supersonic-flow region has negative frequency in a frame co-moving with the fluid, whereas the ray that passes through the system from left to right has positive frequency in the fluid frame.}
\label{fig4}      
\end{figure}

How does the presence of the $k_c$-dependent term in (\ref{dispa}) alter the ray trajectories of Fig.~\ref{fig2}? We numerically solve the Hamilton equations resulting from (\ref{dispa}) (with the positive sign) and obtain the trajectories shown in Fig.~\ref{fig4}.
The qualitative behavior of the rays in Fig.~\ref{fig4} is easily understood from the superluminal character of the dispersion relation (\ref{dispa}). Consider the ray approaching the white hole horizon from the left: the ray moves at the group velocity and is blue-shifted as it nears the horizon; this increases the group velocity until the ray traverses the horizon into the supersonic-flow region. Similarly, the ray being swept towards the white hole horizon from the right is blue-shfted until the group velocity exceeds the speed of the flow, allowing the ray to move away from the horizon. The behavior at the black hole horizon is simply the time reverse of that at the white hole horizon. Thus, the ray in the supersonic-flow region bounces back and forth between the horizons while the ray incident from the left passes through the system.

In Fig.~\ref{fig4} we have actually solved for the two rays using the same positive value of $\omega$, the frequency in the laboratory frame, which is constant throughout the motion. It is then the case that the ray propagating through the system has positive frequency $\omega-vk$ in a frame co-moving with the fluid, whereas the bouncing ray has negative frequency in the fluid frame and does not therefore represent a real phonon. (A real phonon in the supersonic-flow region that is swept towards the white hole horizon has negative $\omega$ and exhibits the same bouncing behavior.) We have chosen the same value of $\omega$ for both rays in order to discuss particle creation. As the ray from the left reaches the horizon geometrical acoustics breaks down, allowing a wave packet centered on this ray to undergo mode conversion wherein it develops a component corresponding to another branch of the dispersion relation having the same conserved value of $\omega$. Thus a wave packet centered on the incoming ray in Fig.~\ref{fig4} will develop a component between the horizons that is centered on the bouncing ray in Fig.~\ref{fig4}, which has negative frequency in the fluid frame. In the quantized theory it is a positive frequency in the fluid frame that is associated with an annihilation operator through a mode expansion such as (\ref{mode}). That fact that a positive-frequency wave packet develops a negative-frequency component implies a mixing of creation and annihilation operators in the quantum theory and this means there is particle creation by the system~\cite{haw75,bir82,bro95b,unr95}. 

The significance of the frequency in the fluid frame is apparent from the inner product
\begin{equation}  \label{norm}
(\phi_1,\phi_2)=i\int\left[\phi^*_1(\partial_t+v\partial_z)\phi_2-\phi_2(\partial_t+v\partial_z)\phi_1^*\right]dz
\end{equation}
which is conserved for solutions of (\ref{unruhmod}), i.e.\ $\partial_t(\phi_1,\phi_2)=0$. The operator $\partial_t+v\partial_z$ is the time derivative in the fluid frame and so is associated with the frequency in this frame. It is easily shown a wave packet $\phi$ made up of plane-wave solutions with positive (negative) frequency in the fluid frame has positive (negative) conserved norm $(\phi,\phi)$.
Since it is the frequency in the fluid frame that is important for particle production and for the inner product (\ref{norm}), the terms positive/negative frequency will hereafter refer to the fluid frame. 

\begin{figure}
\centering
\includegraphics[height=7cm]{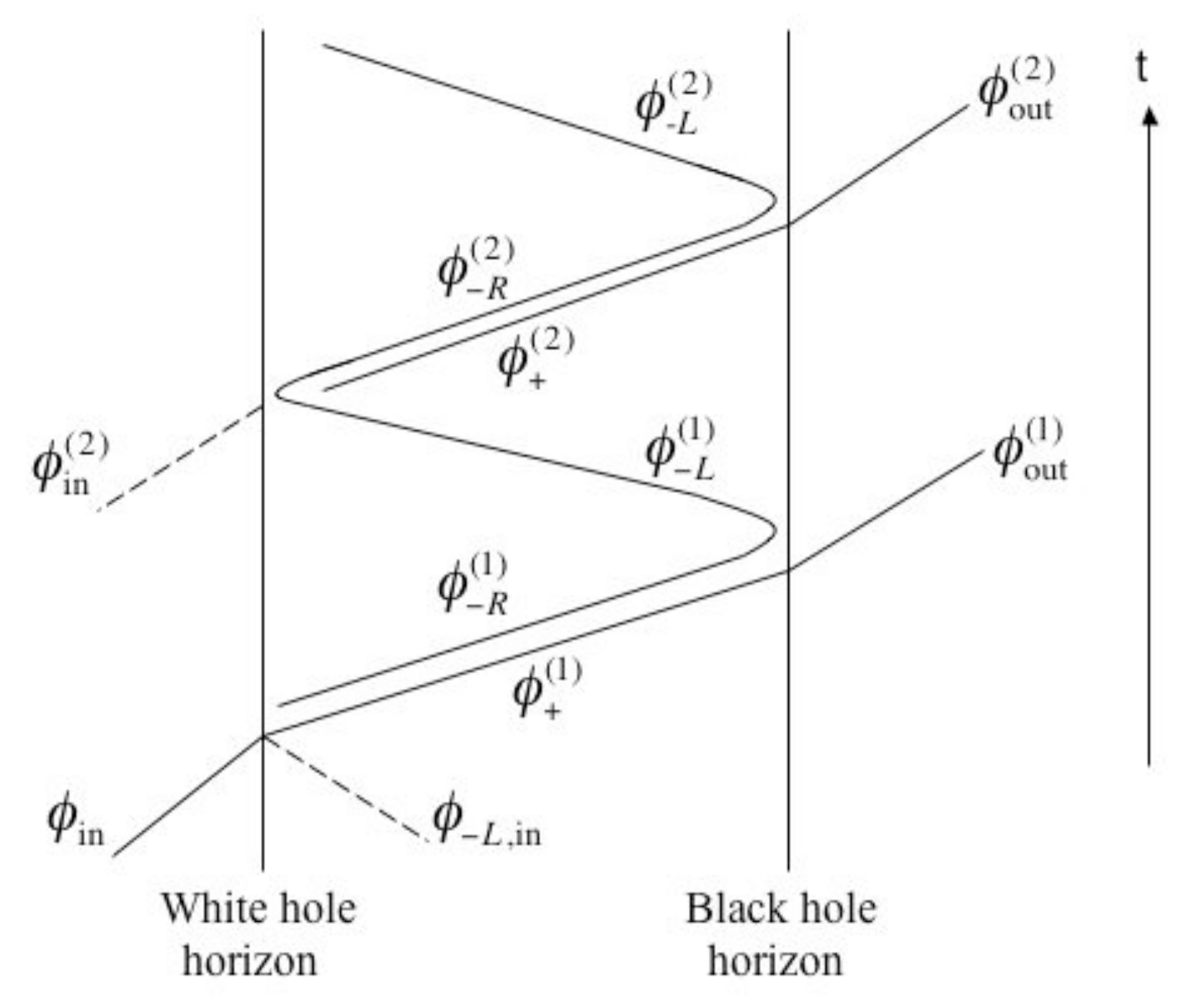}
\caption{Schematic picture of the evolution of a positive-frequency wave packet $\phi_{\mathrm{in}}$. Lines that begin at the horizons denote wave packets created by mode conversion. All wave packets follow one of the ray trajectories of Fig.~\ref{fig4}. The dotted lines indicate modes that are empty in this evolution but that feature in the connection formulae for modes at the horizons.}
\label{fig5}      
\end{figure}

The above considerations lead to the schematic picture of Fig.~\ref{fig5} for the evolution of an incident positive-frequency (and positive-norm) wave packet $\phi_{\mathrm{in}}$. 
At the white hole horizon the packet $\phi_{\mathrm{in}}$ undergoes mode conversion, producing a negative-frequency component $\phi^{(1)}_{-R}$ that propagates to the right together with the positive-frequency part $\phi^{(1)}_+$. When these packets reach the black hole horizon there is further mode conversion and a positive-frequency packet $\phi^{(1)}_\mathrm{out}$ exits the system leaving behind a red-shifted negative-frequency packet $\phi^{(1)}_{-L}$ that propagates to the left towards the white hole horizon. On reaching the white hole horizon $\phi^{(1)}_{-L}$ generates through mode conversion a positive-frequency packet $\phi^{(2)}_+$, which joins the blue-shifted negative-frequency packet $\phi^{(2)}_{-R}$ in propagating towards the black hole horizon. When $\phi^{(2)}_+$ and $\phi^{(2)}_{-R}$ reach the black hole horizon mode conversion occurs again and there is another output of a positive-frequency packet $\phi^{(2)}_\mathrm{out}$, leaving a negative-frequency packet $\phi^{(2)}_{-L}$ between the horizons. The process continues as outlined. Now the total norm of this evolving solution is conserved, so in each instance of mode conversion amounts of positive- and negative-frequency wave packet are created that have equal and opposite norm. Thus with each bounce the norm of the negative-frequency packet between the horizons increases in magnitude; indeed, it increases exponentially since the norms of the wave packets created by mode conversion are in proportion to the norm of the incident packet. The size of the negative frequency packet generated by $\phi_{\mathrm{in}}$ determines the particle production so the number of particles produced increases exponentially with time---this is the black hole laser~\cite{cor99}.

To quantify the above description one can utilize the connection formulae obtained by Corley and Jacobson for modes at the horizons~\cite{cor99}. Let us assume that the slope of the velocity profile is $-\alpha$ at the white hole horizon and $\gamma$ at the black hole horizon. Then the relation between modes with the same conserved value of $\omega$ at the white hole horizon is (see Fig.~\ref{fig5}):
\begin{gather}
\phi_{\omega\mathrm{in}}=\left[2\sinh\left(\pi\omega/\alpha\right)\right]^{-1/2}\left(e^{\pi\omega/2\alpha}\phi^{(1)}_{\omega +}+e^{-\pi\omega/2\alpha}\phi^{(1)}_{\omega -R}\right),  \label{con1} \\[6pt]
\phi_{\omega -L\mathrm{in}}=\left[2\sinh\left(\pi\omega/\alpha\right)\right]^{-1/2}\left(e^{-\pi\omega/2\alpha}\phi^{(1)}_{\omega +}+e^{\pi\omega/2\alpha}\phi^{(1)}_{\omega -R}\right), \label{con2}
\end{gather}
with similar formulae for the other modes that meet at this horizon in Fig.~\ref{fig5}. At the black hole horizon the connection formulae for modes is (see Fig.~\ref{fig5}):
\begin{gather}
\phi^{(1)}_{\omega\mathrm{out}}=\left[2\sinh\left(\pi\omega/\gamma\right)\right]^{-1/2}\left(e^{\pi\omega/2\gamma}\phi^{(1)}_{\omega +}+e^{-\pi\omega/2\gamma}\phi^{(1)}_{\omega -L}\right), \label{con3} \\[6pt]
\phi^{(1)}_{\omega -L}=\left[2\sinh\left(\pi\omega/\gamma\right)\right]^{-1/2}\left(e^{-\pi\omega/2\gamma}\phi^{(1)}_{\omega +}+e^{\pi\omega/2\gamma}\phi^{(1)}_{\omega -L}\right), \label{con4}
\end{gather}
with similar formulae applying for the other modes that interact at this horizon in Fig.~\ref{fig5}. For modes that travel between the horizons ($\phi^{(n)}_{\omega +}$, $\phi^{(n)}_{\omega -R}$ and $\phi^{(n)}_{\omega -L}$) we must in addition specify how their values at the two horizons are related to each other. This can be done using a WKB approximation to propagate the modes from one horizon to the other; the result will be a simple phase relationship between the values at each horizon:
\begin{gather}
\left.\phi^{(n)}_{\omega +}\right|_{\mathrm{BH}}=e^{i\theta_+}\left.\phi^{(n)}_{\omega +}\right|_{\mathrm{WH}}, \label{phase1}  \\[6pt]
\left.\phi^{(n)}_{\omega -R}\right|_{\mathrm{BH}}=e^{-i\theta_-}\left.\phi^{(n)}_{\omega -R}\right|_{\mathrm{WH}}, \label{phase2} \\[6pt]
\left.\phi^{(n)}_{\omega -L}\right|_{\mathrm{WH}}=e^{-i\theta_0}\left.\phi^{(n)}_{\omega -L}\right|_{\mathrm{BH}}, \label{phase3}
\end{gather}
where BH (WH) refers to the black hole (white hole) horizon. Using the above formulae as building blocks one can calculate the evolution of a wave packet and hence compute the number of created particles~\cite{cor99};  in Sec.~\ref{amp} we shall address this task in a different guise. First, a numerical solution for an evolving wave packet will provide an enlightening picture of the process outlined in Fig.~\ref{fig5}.

\section{Numerical results}
\label{numerical}
Let us summarize what we expect to see if we propagate a positive-frequency wave packet towards the white hole horizon. According to Fig.~\ref{fig5} this positive-norm wave packet will plough through the system and emerge out of the black hole horizon ($\phi^{(1)}_{\mathrm{out}}$) somewhat amplified, leaving behind a small negative-norm packet ($\phi^{(1)}_{-L}$) such that the total norm is conserved. After a period of time equal to that required by the negative-frequency ray in Fig.~\ref{fig4} to propagate from the black hole horizon to the white hole horizon and back again we expect another, smaller, positive-norm packet ($\phi^{(2)}_{\mathrm{out}}$) to emerge from the system. At regular intervals there should be further outputs of positive-norm packets ($\phi^{(3)}_{\mathrm{out}}$, etc.) and a build-up of the negative-norm packet between the horizons. The increasing amplitude of the negative-norm packet will cause an increase in the size of the positive-norm outputs produced by mode conversion, and this increase should in fact be exponential.

\begin{figure}
\centering
\includegraphics[height=8cm]{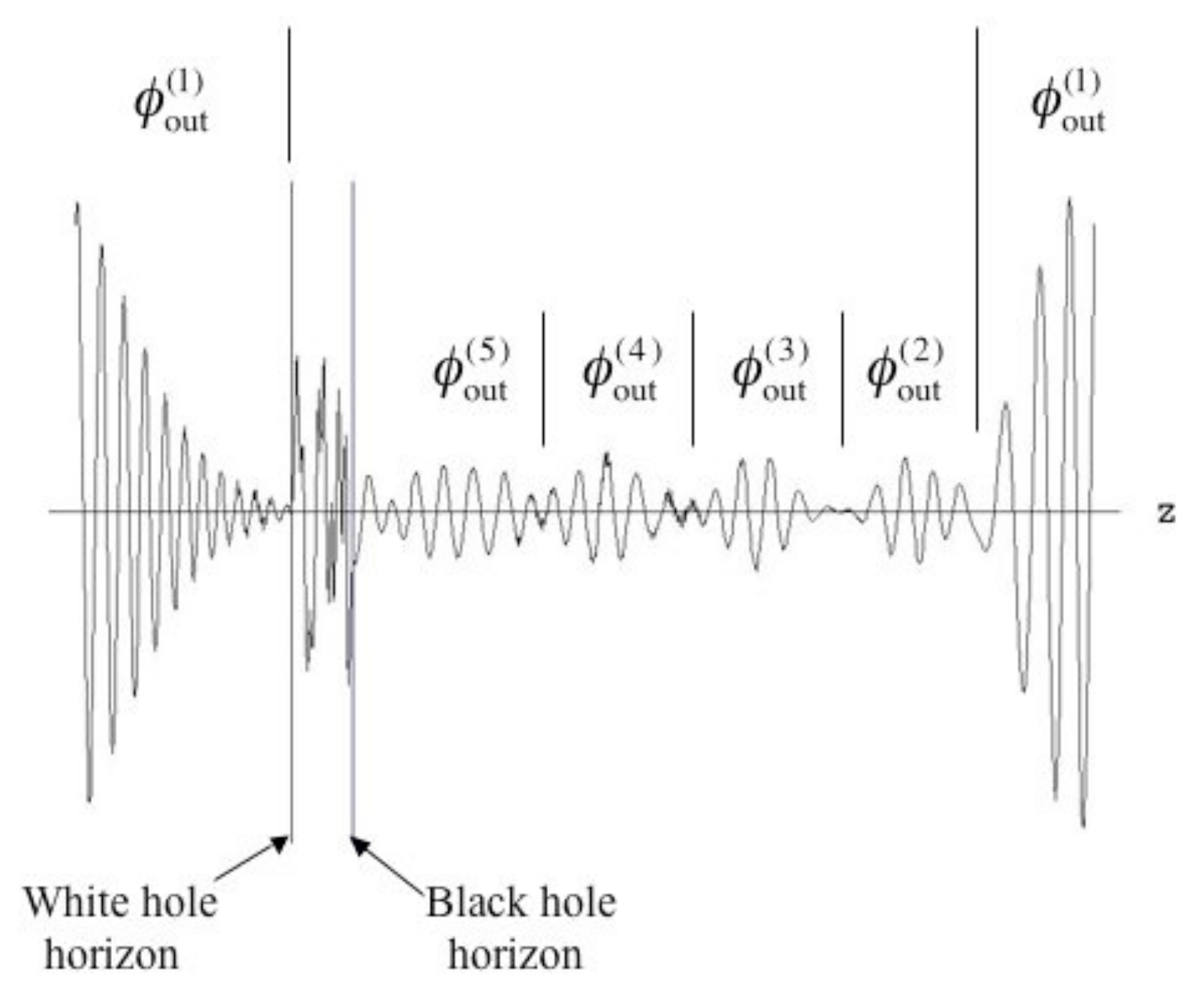}
\caption{Numerical solution for a positive-frequency wave packet that has entered the horizon system from the left. The packet is centered on the positive-frequency ray in Fig.~\ref{fig4} so it has propagated through the system ($\phi^{(1)}_{\mathrm{out}}$) and begun to emerge on the left-hand side due to periodic boundary conditions. A sequence of positive-frequency wave packets are seen to follow $\phi^{(1)}_{\mathrm{out}}$, in line with expectations based on Fig.~\ref{fig5}.}
\label{fig6}      
\end{figure}

In Fig.~\ref{fig6} we show the result of a numerical evolution of a wave packet centered on the positive-frequency ray in Fig.~\ref{fig4}.
Periodic boundary conditions are in force in Fig.~\ref{fig6} and the incident packet has ploughed through the system ($\phi^{(1)}_{\mathrm{out}}$) and is circling round on the left-hand side where it is about to hit the white hole horizon a second time. The successive outputs of positive-norm packets ($\phi^{(2)}_{\mathrm{out}}$, etc.) are clearly visible; five separate emissions are distinguishable, corresponding to the negative-norm packet bouncing off the black hole horizon five times. As the total amount of  positive-norm packet increases with each output it is balanced by a growing amount of negative-norm packet between the horizons, leaving the total norm unchanged.

The distance between the horizons has been set quite small relative to the length of the incident wave packet in order to increase the number of bounces that occur before $\phi^{(1)}_{\mathrm{out}}$ collides with the white hole horizon. This makes it difficult to see clearly the motion of the negative-frequency wave packet that is confined to the supersonic-flow region. Nevertheless alternate red- and blue-shifting is observed between the horizons as the solution evolves and the time between successive positive-norm outputs is equal to the time between visits to the black hole horizon of the ray on which the negative-frequency packet is centered (see Fig.~\ref{fig4}).

Although the last output $\phi^{(5)}_{\mathrm{out}}$ in Fig.~\ref{fig6} seems to be somewhat larger than $\phi^{(2)}_{\mathrm{out}}$ there is no clear sign of an exponential increase in the size of the output packets after this number of emissions. The analytic formulae for the evolution show that about ten outputs would be required in this simulation before there is a significant increase in their size, as we shall see in the next Section.

Before we turn to an analytic treatment of the black hole laser and the calculation of the particle production, we briefly mention the equally remarkable results when the field $\phi$ is fermionic rather than bosonic~\cite{cor99}. The conserved norm for a fermionic field is positive definite, so both positive- and negative-frequency packets have positive norm. From this fact alone follows the evolution of a positive-frequency fermionic wave packet. When the initial wave packet ploughs through the system as in Fig.~\ref{fig6} it is diminished in amplitude in order to conserve the total norm, since the negative-frequency packet left behind also has positive norm. As subsequent positive-frequency packets are emitted conservation of norm requires the amount of negative frequency packet between the horizons to decrease rather than increase as in the bosonic case. In the limit of infinite time the amount of negative frequency packet goes to zero and therefore so does the number of particles produced---there is complete suppression of the Hawking effect. This can be understood as a filling up of all the allowed fermionic states between the horizons by the Hawking particles created in this region~\cite{cor99}.

\section{Black hole amplifier}
\label{amp}
The Bogoliubov transformation that underlies particle creation by a black hole is similar to that which describes the action of a parametric amplifier or a phase-conjugating mirror in quantum optics~\cite{she84,leo03}. An optical device will transform a set of incident quantum light modes, described by annihilation operators $\{\hat{a}_1,\dots,\hat{a}_n\}$ and their Hermitian conjugates $\{\hat{a}_1^\dagger,\dots,\hat{a}_n^\dagger\}$, into outgoing modes, described by operators $\{\hat{a}{'}_{\!\!1},\dots,\hat{a}{'}_{\!\!n}\}$ and $\{\hat{a}{'}_{\!\!1}^\dagger,\dots,\hat{a}{'}_{\!\!n}^\dagger\}$. If the device is linear this transformation is given by the scattering matrix $S$:
 \begin{equation} \label{Sdef}
    \begin{pmatrix}  \hat{a}{'}_{\!\!1}  \\ \vdots \\ \hat{a}{'}_{\!\!n} \\  \hat{a}{'}_{\!\!1}^\dagger  \\ \vdots \\ 
  \hat{a}{'}_{\!\!n}^\dagger \end{pmatrix} =S    \begin{pmatrix}  \hat{a}_1  \\ \vdots \\ \hat{a}_n \\  \hat{a}_1^\dagger  \\ \vdots \\ \hat{a}_n^\dagger \end{pmatrix}.
\end{equation}
The $S$-matrix is constrained by the requirement that the creation and annihilation operators satisfy the usual commutation relations; this imposes the quasi-unitarity 
condition
\begin{equation} \label{quasi}
SGS^\dagger=G, \qquad G=   \begin{pmatrix} 
      I & 0 \\
      0 & -I \\
   \end{pmatrix}.
\end{equation}
Matrices $S$ satisfying (\ref{quasi}) form the group $U(n,n)$~\cite{cor}.
An example of such a matrix is
\begin{equation} \label{scat}
S= \begin{pmatrix} 
      \cosh\xi\  & 0 & 0 & \ \sinh\xi \\
      0 & \ \cosh\xi\  & \ \sinh\xi\  & 0 \\
      0 & \sinh\xi & \cosh\xi & 0 \\
     \sinh\xi & 0 & 0 & \cosh\xi \\
    \end{pmatrix},
\end{equation}
which is the scattering matrix for a parametric amplifier or a phase-conjugating mirror~\cite{she84,leo03}. In this case there are two input and output modes and the device creates or annihilates pairs of photons. The energy for the pair production or the reservoir for pair annihilation is provided by the pump process of the amplifier, which determines the parameter $\xi$. For a black hole we may define $\xi$ by
\begin{equation} \label{xi}
\tanh\xi=e^{-\pi\omega/\gamma},
\end{equation}
where $\gamma$ is the surface gravity at the horizon and $\omega$ is the frequency of the quantum modes under consideration. Then (\ref{scat}) gives the Bogoliubov transformation for the Hawking effect~\cite{bir82,bro95b}, which we write concisely as
\begin{equation} \label{bogbh}
    \begin{pmatrix}  \hat{a}{'}_{\!\!1}  \\   \hat{a}{'}_{\!\!2}^\dagger \end{pmatrix} = 
    \begin{pmatrix} 
      \cosh\xi\  & \ \sinh\xi \\
     \sinh\xi & \cosh\xi \\
    \end{pmatrix}
     \begin{pmatrix}  \hat{a}_1  \\   \hat{a}_2^\dagger \end{pmatrix} .
\end{equation}

For a moving fluid, (\ref{xi}) and (\ref{bogbh}) describe phonon creation by a black hole horizon, where $\gamma$ denotes the slope of the velocity profile $v(z)$ at the horizon. A white hole horizon in the fluid, where the slope of $v(z)$ is $-\alpha$, is the time reverse of a black hole horizon. The ``surface gravity" is $\alpha$ and the roles of input and output modes are  reversed. Thus, defining a parameter $\zeta$ by
\begin{equation} \label{zeta}
\tanh\zeta=e^{-\pi\omega/\alpha}
\end{equation}
the white hole horizon generates the Bogoliubov transformation
\begin{equation} \label{bogwh}
    \begin{pmatrix}  \hat{a}{'}_{\!\!1}  \\   \hat{a}{'}_{\!\!2}^\dagger \end{pmatrix} = 
    \begin{pmatrix} 
      \cosh\zeta\ \  & \ -\sinh\zeta \\
     -\sinh\zeta & \ \ \cosh\zeta \\
    \end{pmatrix}
     \begin{pmatrix}  \hat{a}_1  \\   \hat{a}_2^\dagger \end{pmatrix} .
\end{equation}

\begin{figure}
\centering
\includegraphics[height=6cm]{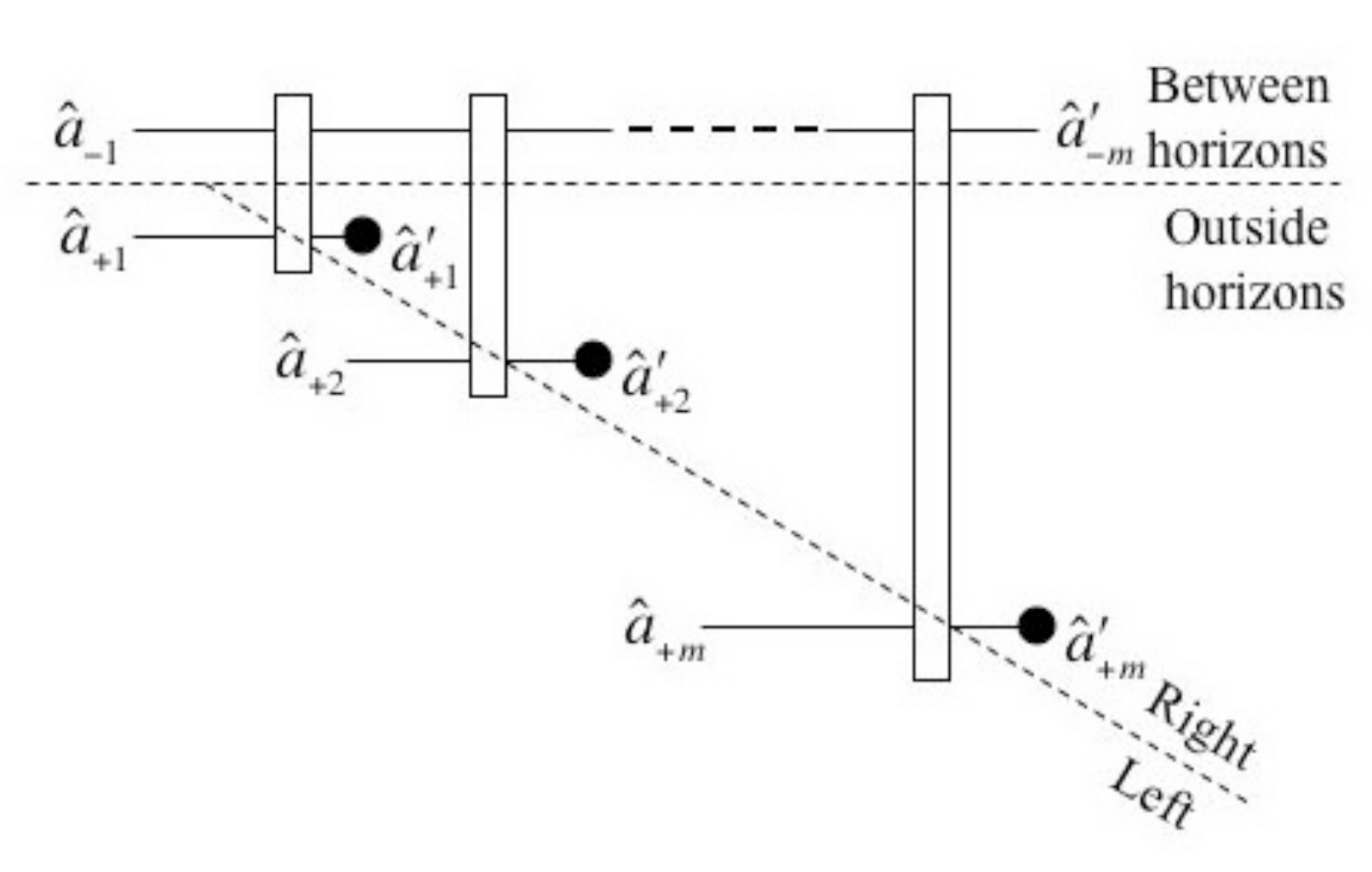}
\caption{The black hole laser as a network of amplifiers. The mode operators with negative subscripts correspond to the Hawking radiation trapped between the horizons, while those with positive subscripts describe the Hawking radiation present outside the horizons. An input consisting of a mode on the left of the black hole laser and a mode between the horizons produces an output mode on the right of the black hole laser and another output mode between the horizons, as in Fig.~\ref{fig5}. The process repeats, with the output mode between the horizons acting as an input mode for the next cycle, i.e.\ $\hat{a}{'}_{\!\!-n}=\hat{a}_{-(n+1)}$.}
\label{BHamp}      
\end{figure}

In the black hole laser the particle production by the pair of horizons is not given by (\ref{bogbh}) and (\ref{bogwh}) because the horizons interact due to the superluminal dispersion. We can represent the black hole laser as a quantum-optical device by expressing the amplification process of Fig.~\ref{fig5} in terms of the annihilation operators of the modes, and using a more convenient notation (see Fig.~\ref{BHamp}). In the $n$th amplification step we have a pair of input modes $\hat{a}_{+n}$ and $\hat{a}_{-n}$ at the white hole horizon that generates an output mode $\hat{a}{'}_{\!\!+n}$ on the right-hand side of the black hole laser as well as a mode $\hat{a}{'}_{\!\!-n}$ that is trapped between the horizons where it serves as a further input to the device, i.e.
 \begin{equation} \label{outin}
 \hat{a}{'}_{\!\!-n}=\hat{a}_{-(n+1)}.
 \end{equation}
 The picture is that of a network of amplifiers, as outlined in Fig.~\ref{BHamp}. 

The output of the black hole laser is completely specified by the matrix $B$ that effects the Bogoliubov transformation for each step in Fig.~\ref{BHamp}:
\begin{equation} \label{step}
    \begin{pmatrix}  \hat{a}{'}_{\!\!-n}  \\   \hat{a}{'}_{\!\!+n}^\dagger \end{pmatrix} = B
     \begin{pmatrix}  \hat{a}_{-n}  \\   \hat{a}_{+n}^\dagger \end{pmatrix} .
\end{equation}
There are four distinct processes incorporated in the matrix $B$, allowing it to be expressed as
\begin{equation} \label{B}
B=B_4B_3B_2B_1.
\end{equation}
First there is the pair production by the white hole horizon, described by (\ref{bogwh}); we therefore have
\begin{equation} \label{B1}
    B_1=    \begin{pmatrix} 
      \cosh\zeta\ \  & \ -\sinh\zeta \\
     -\sinh\zeta & \ \ \cosh\zeta \\
    \end{pmatrix}.
\end{equation}
 Next, the two modes produced by the white hole horizon propagate to the black hole horizon, as in Fig.~\ref{fig5}; this imparts separate phases $\theta_-$ and $\theta_+$ to each mode operator, so we write 
\begin{equation} \label{B2}
    B_2=    \begin{pmatrix} 
      e^{-i\theta_-}  & 0 \\
     0 &   e^{i\theta_+} \\
    \end{pmatrix}.
\end{equation}
At the black hole horizon the modes are transformed according to (\ref{bogbh}), giving
\begin{equation} \label{B3}
    B_3=    \begin{pmatrix} 
      \cosh\xi\ & \ \sinh\xi \\
     \sinh\xi & \ \cosh\xi \\
    \end{pmatrix}.
\end{equation}
Finally the mode trapped between the the horizons propagates to the white hole horizon, acquiring a phase $\theta_0$:
 \begin{equation} \label{B4}
    B_4=    \begin{pmatrix} 
      e^{-i\theta_0}  & 0 \\
     0 &   1 \\
    \end{pmatrix}.
\end{equation}

The particle content after $m$ amplification steps is now easily calculated from (\ref{step}) and (\ref{outin}), using the value (\ref{B})--(\ref{B4}) for $B$. It is worthwhile to consider the restrictions on the $2\times 2$ matrix $B$ resulting from the fact that it determines a $U(2,2)$ scattering matrix $S$ (see (\ref{step}) and (\ref{Sdef})). Constructing $S$ from $B$ and imposing (\ref{quasi}) one finds that the components of $B$ must satisfy
\begin{equation} \label{Bcpts}
|B_{11}|^2-|B_{12}|^2=1, \qquad |B_{21}|^2-|B_{22}|^2=1, \qquad B_{11}^*B_{21}-B_{12}^*B_{22}=0,
\end{equation} 
which is the statement that $B$ is a $U(1,1)$ matrix. The constraints (\ref{Bcpts}) impose the following structure on $B$:
\begin{gather} 
  B=    e^{i\psi}\begin{pmatrix} 
      \mu\  & \nu^* \\
     \nu\  &   \mu^* \\
    \end{pmatrix}, \quad\text{$\psi$ real},  \label{Bform} \\[6pt]
     |\mu|^2-|\nu|^2=1.  \label{munu}
\end{gather}
Using (\ref{step}), (\ref{outin}) and (\ref{Bform}) we find after $m$ cycles that $\hat{a}{'}_{\!\!-m} $ for the mode trapped between the horizons and  $\hat{a}{'}_{\!\!+m}^\dagger$ for the mode outside the horizons are given by
\begin{gather}
\hat{a}{'}_{\!\!-m} =e^{im\psi}\mu^m\hat{a}_{-1}+\nu^*\sum_{l=1}^me^{i(m-l+1)\psi}\mu^{m-l}  \hat{a}_{+l}^\dagger, \\[6pt]
\hat{a}{'}_{\!\!+m}^\dagger =e^{im\psi}\nu\mu^{m-1}\hat{a}_{-1}+|\nu |^2\sum_{l=1}^{m-1}e^{i(m-l+1)\psi}\mu^{m-l-1}\hat{a}_{+l}^\dagger+e^{i\psi}\mu^*\hat{a}_{+m}^\dagger.
\end{gather}
The condition that we have vacuum before the formation of the black hole laser gives
\[
\hat{a}_{-1}|0\rangle=0, \qquad \hat{a}_{+l}|0\rangle=0, \quad l=1,\dots m.
\]
We may write the particle content for the two output modes in terms of $|\mu|^2$ because of (\ref{munu}). The particle number for the trapped mode after $m$ cycles is
\begin{eqnarray}
\langle\hat{N}_{-m}\rangle&=&\langle \hat{a}{'}_{\!\!-m}^\dagger\hat{a}{'}_{\!\!-m}\rangle \nonumber \\[6pt]
&=&|\nu|^2\sum_{l=1}^{m}|\mu|^{2(m-l)}=|\nu|^2\frac{|\mu|^{2m}-1}{|\mu|^2-1} \nonumber \\[6pt]
&=&|\mu|^{2m}-1, \label{N-}
\end{eqnarray}
whereas the number of particles outside the horizons is
\begin{eqnarray}
\langle\hat{N}_{+m}\rangle&=&\langle \hat{a}{'}_{\!\!+m}^\dagger\hat{a}{'}_{\!\!+m}\rangle  \nonumber \\[4pt]
&=&|\nu|^2|\mu|^{2(m-1)} \nonumber \\[4pt]
&=&|\mu|^{2m}\left(1-|\mu|^{-2}\right). \label{N+}
\end{eqnarray}
The explicit form of $|\mu|^2$ is found from (\ref{B})--(\ref{B4}) and (\ref{Bform}) to be
\begin{equation} \label{musquared}
|\mu|^2=\frac{1}{2}\Big[1+\cosh (2\xi)\cosh(2\zeta)-\cos(\theta_++\theta_-)\sinh(2\xi)\sinh(2\zeta)\Big].
\end{equation}
Note that the phase $\theta_0$ does not appear in (\ref{musquared}) and thus has no effect on the particle content; this is because $\theta_0$ is acquired by a mode traveling  to the white hole horizon where it interacts with a mode from outside the device, and the latter has no phase sensitivity. The combination $\theta_++\theta_-$, however, is the phase difference between the two modes that propagate across the device from the white hole horizon and this affects how these modes interact at the black hole horizon. We see from (\ref{N-})--(\ref{musquared}) that for frequencies such that  $\cos(\theta_++\theta_-)=1$ the particle production is maximally suppressed; indeed if the horizons are symmetric ($\xi=\zeta$) there is no production at this frequency. On the other hand maximal amplification of the Hawking radiation occurs at frequencies for which $\cos(\theta_++\theta_-)=0$, giving a WKB-type condition
\[
\theta_++\theta_-=2\pi\left(n+\frac{1}{2}\right), \qquad \text{$n$ an integer.}
\]

\begin{figure}
\centering
\includegraphics[height=4cm]{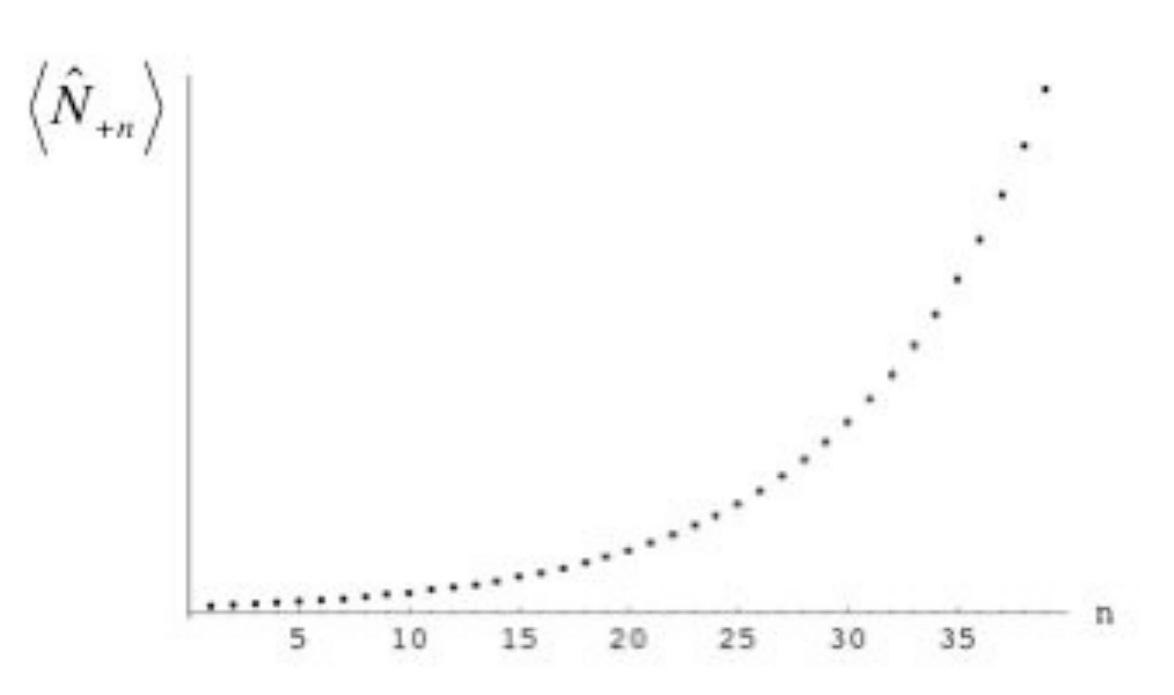}
\caption{The number of Hawking particles outside the horizons versus the number of amplification cycles,  as given by formula (\ref{N+}).}
\label{Nfig}      
\end{figure}

In Fig.~\ref{Nfig} we use formula (\ref{N+}) to plot the particle number outside the horizons as a function of the number of amplification cycles.
The slopes of $v(z)$ at the horizons are those used for the numerical example in Sec.~\ref{numerical} and the frequency chosen is that on which the incident pulse in Fig~\ref{fig6} is centered. A WKB approximation is used to estimate the phases $\theta_+$ and $\theta_-$. Fig.~\ref{Nfig} shows the consequence, in terms of Hawking radiation, of the wave-packet evolution in Fig.~\ref{fig6}. As discussed in Sec.~\ref{numerical}, the stream of positive-frequency pulses emerging from the black hole horizon in Fig.~\ref{fig6} implies a growth in particle production, as in Fig.~\ref{Nfig}. The solution domain in Fig.~\ref{fig6} is only large enough to contain the result of five amplification cycles. In this time the positive-frequency pulses trailing $\phi^{(1)}_{\text{out}}$ are of approximately the same size; this implies an approximately linear increase in the particle number, as is the case in Fig.~\ref{Nfig} with $n\leq 5$. In order for the exponential nature of the increase in particle number to be clearly seen in a wave-packet evolution, one would require at least ten positive-frequency  outputs in the evolution, as one sees from Fig.~\ref{Nfig}.

In summary, the black hole laser is a dramatic quantum-vacuum effect and provides an interesting example of the interplay of Hawking radiation and nonlinear dispersion. Any cosmological significance of this process remains highly speculative~\cite{cor99}, but in contrast a laboratory example of black hole lasing is not nearly so far-fetched. Bose--Einstein condensates have the properties required for the construction of a sonic black hole laser, however formidable the practical difficulties may be, and this is not the only current possibility~\cite{vol05}. As the study of black hole analogues continues the prospect of exponentially amplified Hawking radiation is tantalizing for both theorists and experimentalists.

\end{document}